\documentclass{kluwer}    

\newcommand{\gapprox}{\lower.4ex\hbox{$\;\buildrel >\over{\scriptstyle\sim}\;$}}
\newcommand{\lapprox}{\lower.4ex\hbox{$\;\buildrel 
<\over{\scriptstyle\sim}\;$}}

\usepackage{epsfig}

\begin{document}
\begin{article}
\begin{opening}
\title{         On the Photometric Accuracy of RHESSI
		Imaging and Spectrosocopy 		}

\author{Markus J. \surname{Aschwanden}$^1$,
                Thomas R. \surname{Metcalf}$^1$, 
		S\"am \surname{Krucker}$^2$,  
		Jun \surname{Sato}$^3$,
		Andrew J. \surname{Conway}$^4$,
                Gordon J. \surname{Hurford}$^2$, and
		Ed, J. \surname{Schmahl}$^{5,6}$}

\runningauthor{ASCHWANDEN ET AL.}

\runningtitle{RHESSI PHOTOMETRY}

\institute{ $^1$) Lockheed Martin Advanced Technology Center,
                  Solar \& Astrophysics Laboratory,
                  Dept. L9-41, Bldg.252,
                  3251 Hanover St.,
                  Palo Alto, CA 94304, USA;
                  e-mail: aschwanden@lmsal.com, \\
            $^2$) Space Sciences Laboratory,
		  University of California at Berkeley,
		  Berkeley, CA 94720-7450;\\
	    $^3$) Dept. of Physics,
		  Montana State University,
		  Bozeman, MO 59717;\\
	    $^4$) Dept. Physics and Astronomy,
		  The Open University,
		  Milton Keynes MK7 6AA, UK;\\
	    $^5$) Solar \& Astrophysics Laboratory,
		  NASA Goddard Space Flight Center,
		  Greenbelt, MD 20770, USA;\\
	    $^6$) Astronomy Dept,
		  University of Maryland,
		  College Park, MD 20742.}

\date{2003/Aug/14}

\begin{abstract}
We compare the photometric accuracy of spectra and images
in flares observed with the {\sl Ramaty 
High Energy Solar Spectroscopic Imager (RHESSI)} spacecraft. 
We test the accuracy of the photometry by comparing the 
photon fluxes obtained in different energy ranges from the
spectral-fitting software SPEX with those fluxes contained 
in the images reconstructed with the {\sl Clean, MEM,
MEM-Vis, Pixon,} and {\sl Forward-fit} algorithms.
We quantify also the background fluxes, the fidelity 
of source geometries, and spatial spectra reconstructed
with the five image reconstruction algorithms. We investigate
the effects of grid selection, pixel size, field-of-view,
and time intervals on the quality of image reconstruction.
The detailed parameters and statistics are provided in an
accompanying CD-ROM and web page.
We find that {\sl Forward-fit}, {\sl Pixon}, and {\sl Clean}
have a robust convergence behavior and a photometric
accuracy in the order of a few percents, while {\sl MEM}
does not converge optimally for large degrees
of freedom (for large field-of-views and/or small pixel sizes), 
and {\sl MEM-Vis} suffers in the case of time-variable
sources. This comparative study 
documents the current status of the {\sl RHESSI} spectral and 
imaging software, one year after launch. 
\end{abstract}


\end{opening}

\section{              	INTRODUCTION                    }

The {\sl Reuven Ramaty High Energy Solar Spectroscopic Imager (RHESSI)}
was launched on 2002 Feb 5 and operated successfully through the first
year, recording a total of $\approx 7500$ solar flares. 
At this time, after one year of the mission, most of the software
has matured to be both reliable and accurate, and instrumental parameters
have been improved from in-flight data to great precision. Thus, it is
timely to conduct a quantitative comparison of the performance of
the various available imaging algorithms, to compare the spatial and spectral
features in the reconstructed images, and to cross-calibrate
the photon fluxes extracted from images against those from the
(non-imaging) spectral fitting software. Such a quantitative comparison
will also be useful to provide estimates of uncertainties and systematic
errors in spectral fitting and spectral inversions, as well as to  
serve as a reference for monitoring future changes in the instrumental 
parameters and software developments. All image reconstructions are
performed here with the software version available in March 2003.

The {\sl RHESSI} spacecraft and its mission are described in the overview
article of Lin et al. (2002), the {\sl RHESSI} spectrometer in Smith et al.
(2002), and the {\sl RHESSI} imaging concepts in Hurford et al. (2002).
Further technical articles that may be of interest in the context of this
study are on {\sl RHESSI} data analysis software (Schwartz et al. 2002),
on reconstruction of {\sl RHESSI} images with forward-fitting (Aschwanden
et al. 2002a), and on pixon image reconstruction (Metcalf et al. 1996). 
We choose for this comparative study as main event the flare of 2002 Feb 20,
11:06 UT,
because this event was already subject of various analyses in a number
of previous publications (Krucker \& Lin 2002, Sui et al. 2002, 
Vilmer et al. 2002, Brown et al. 2002, Aschwanden et al. 2002b, 
Kontar et al. 2002), as well as the two additional flares of
2002 May 18, 19:15 UT, and 2002 Jul 23, 00:30 UT.

\section{ 		RESULTS OF DATA ANALYSIS			}

The details of the data analysis are provided on the accompanying CD-ROM
(or on the web-page {\sl http://www.lmsal.com/\~{}aschwand/eprints/} 
{\sl 2003$\_$photo/index.html}).
Here we summarize only the main results of this analysis, which includes a
detailed parametric study of the 2002 Feb 20 flare (\S A1-A11 on CD-ROM), 
as well as additional tests on flares with different source morphologies and 
flare positions, i.e., the single-source flare from  2002 May 18 (\S B1 on CD-ROM) 
and the complex-source X-ray flare of 2002 Jul 23 (\S C1 on CD-ROM). 

\subsection{ 		Spectral Photometry (\S A1 on CD-ROM)			}

The photometry can be obtained from two different software codes in the
{\sl RHESSI SolarSoftWare (SSW)}, either from the spectral fitting package 
{\sl SPEX} (Schwartz et al. 2002; Smith et al. 2002), or from the {\sl RHESSI}
image reconstruction algorithms. Both codes use the same input in form of
time-tagged photon event lists, but the {\sl RESSI} response function is applied
by independent modules to obtain photon counts in a given energy range. A first
basic photometry test that verifies the consistency of the software is therefore
to compare the photon counts of the {\sl SPEX} spectra in different energy ranges 
with those obtained from images in the same energy ranges. The time profile of
the flare, the photon spectrum, the photon count rates and pre-flare 
background in 8 energy ranges are all shown in \S A1 on the CD-ROM. The pre-flare 
background amounts to a few percent of the peak flux (at energies $\lapprox 40$ keV). 

\begin{table}
\footnotesize
\baselineskip10pt
\begin{tabular}{lrrrrrrr}
\hline
Energy&Forward 	  &Pixon &    Clean &      MEM &    MEM-Vis &    Average \\
range &flux ratio &flux ratio &flux ratio &flux ratio &flux ratio &flux ratio \\
\hline
10-12 &   403.41 &   378.50 &   369.52 &   358.38 &   303.28 & 362.62$\pm$ 37.09 \\ 
15-17 &    56.72 &    55.76 &    55.20 &    50.79 &    50.16 &  53.72$\pm$  3.02 \\ 
20-22 &    18.99 &    18.96 &    19.13 &    17.21 &    17.70 &  18.40$\pm$  0.88 \\ 
25-29 &    15.76 &    15.38 &    16.22 &    14.16 &    14.84 &  15.27$\pm$  0.80 \\ 
30-34 &     8.56 &     8.25 &     8.54 &     7.85 &     7.29 &   8.10$\pm$  0.54 \\ 
35-39 &     5.19 &     5.33 &     5.38 &     9.60 &     4.74 &   6.05$\pm$  2.00 \\ 
40-50 &     7.01 &     6.80 &     7.31 &     6.33 &     6.59 &   6.81$\pm$  0.38 \\ 
50-60 &     3.49 &     3.25 &     3.51 &     3.14 &     2.91 &   3.26$\pm$  0.25 \\ 
\hline
10-12 &   0.98 &   0.92 &   0.90 &   0.87 &   0.74 &   0.88$\pm$  0.09 \\ 
15-17 &   1.02 &   1.01 &   0.99 &   0.92 &   0.90 &   0.97$\pm$  0.05 \\ 
20-22 &   1.03 &   1.03 &   1.04 &   0.93 &   0.96 &   1.00$\pm$  0.05 \\ 
25-29 &   1.05 &   1.02 &   1.08 &   0.94 &   0.98 &   1.01$\pm$  0.05 \\ 
30-34 &   1.04 &   1.00 &   1.04 &   0.95 &   0.89 &   0.99$\pm$  0.07 \\ 
35-39 &   1.02 &   1.04 &   1.06 &   1.88 &   0.93 &   1.19$\pm$  0.39 \\ 
40-50 &   1.11 &   1.08 &   1.16 &   1.01 &   1.05 &   1.08$\pm$  0.06 \\ 
50-60 &   0.98 &   0.91 &   0.99 &   0.88 &   0.82 &   0.92$\pm$  0.07 \\ 
\hline\
Average &   1.03 &   1.00 &   1.03 &   1.05 &   0.91 & \\
        & $\pm$  0.04 & $\pm$  0.06 & $\pm$  0.08 & $\pm$  0.34 & $\pm$  0.10 & \\
\hline
\end{tabular}
\caption[]{Ratios of image-integrated photon fluxes  
	to the total spectral fluxes from SPEX. The last column shows
	the means and standard deviations of all 5 imaging methods.}
\end{table}

\subsection{ 		Image Photometry (\S A2 on CD-ROM)			}

We conduct a photometric test by comparing the photon rates (in units of {\sl photons s$^{-1}$
cm$^{-2}$}) integrated over each image with those obtained from the total flux spectra
calculated with {\sl SPEX} in identical energy bins. We find that these total fluxes
are well conserved in most of the imaging reconstruction algorithms (see ratios in Table 1), 
they agree with {\sl SPEX} within a few percents for all 5 imaging algorithms ({\sl Forward-fit, 
Pixon, Clean, MEM, MEM-Vis}). 
The photometric accuracy seems to be uniform in different energy ranges (last column in Table 1).
This test essentially verifies that the flux normalization is correctly and consistently
implemented in the different software modules of each imaging algorithm.

\begin{figure} 
\includegraphics[bb=50 50 609 350,width=\textwidth]{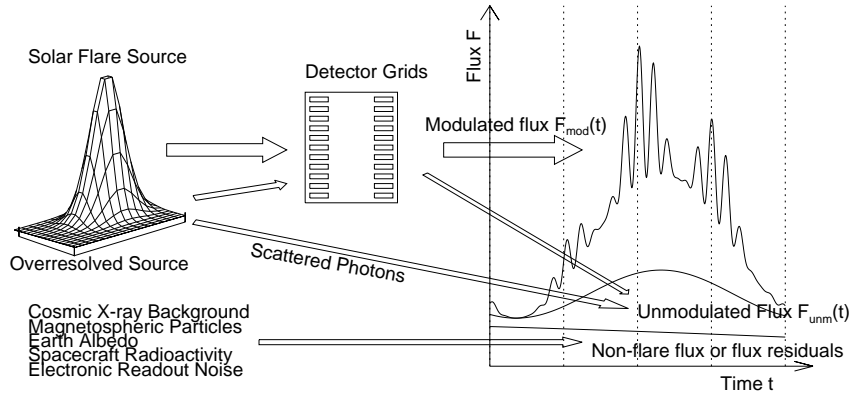}
\caption{Schematic cartoon of solar and non-solar sources that produce
modulated flux $F_{mod}(t)$, unmodulated flux $F_{unm}(t)$, and non-flare
sources $F_{back}(t)$ in the time domain of the modulation time profiles.}
\end{figure}

\subsection{ 		Background Components (\S A3 on CD-ROM)			}

The imaged flux consists of at least 3 different components that can be distinguished in the
image reconstruction process: (1) modulated flux $F_{mod}(t)$ that is modulated by the rotating
grids and originates from resolved flare sources, (2) unmodulated flux $F_{unm}(t)$, which shows no
grid modulation but varies proportionally to the smoothed flare flux, probably produced by
scattered flare photons or by overresolved sources that are larger than the pitch of the
coarsest grid, and (3) non-flare background flux $F_{back}(t)$ that may originate
from cosmic X-ray background, magnetospheric particles, Earth albedo, spacecraft radioactivity,
or electronic readout noise (Fig.~1). Background modeling can be performed with the 
{\sl Pixon} and {\sl Forward-Fit} algorithms. We find an (unexplained) unmodulated flux  
in the order of $\approx 15\%$, in excess of the preflare background rate ($\approx 1\%$). Although
the thin attenuator was inserted during this flare, it has probably no effect on the
unmodulated flux because it is located after the grids.  

\begin{figure} 
\includegraphics[bb=30 360 550 750,width=\textwidth]{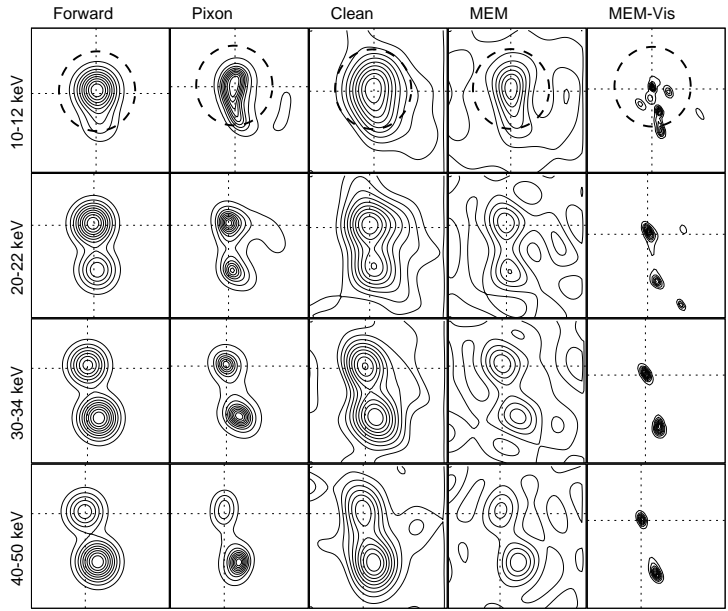}
\caption{Image comparisons of {\sl Forward-fit, Pixon, Clean, MEM,} and {MEM-Vis}
images reconstructed in 4 energy bins. The contour levels are at 10\%,
20\%, ..., 90\% of the maximum flux. The location of the northern flux maximum is
indicated with a dotted cross-hair. The dotted circle has a radius of 18", which
is used for halo/core flux ratio measurements.}
\end{figure}

\subsection{ 		Positional Comparisons (\S A4 on CD-ROM)			}

Images reconstructed in 4 energy bands with all 5 algorithms are shown in Fig. 2.
The source morphology is a simple double footpoint geometry, which is consistently
reproduced by all 5 algorithms. MEM-Vis in its current implementation does not take
corrections for time variability into account, and thus seems to have a compromised
image quality for this case (integrated over a time interval of 40 s). 
We measure the centroid location $(x_N, y_N)$ 
of the northern flare source with sub-pixel accuracy by parabolic interpolation at 
the source peak. Averaging these source positions among the different image algorithms 
we find a standard deviation in the order of 0.5 pixels (with pixel size of 1"), which
is close to the accuracy expected from simulations for the corresponding count rates.
Only MEM-Vis shows occasionally larger deviations, very likely to be a side-effect of
the compromised image quality.

\subsection{ 		Amplitude and Width Comparisons (\S A5 on CD-ROM)			}

The reconstructed source widths $w$ vary to a larger degree between
individual imaging algorithms. Compared with the spatial resolution of
the finest used grid ($w_{res}\approx 11.8"$ for grid\#4), we find that
the width ratio of the reconstructed nonthermal sources at 25 keV is
$w/w_{res} = 1.2$ for {\sl Forward-Fit},
$w/w_{res} = 0.7$ for {\sl Pixon},
$w/w_{res} = 1.7$ for {\sl Clean},
$w/w_{res} = 3.3$ for {\sl MEM}, and
$w/w_{res} = 0.2$ for {\sl MEM-Vis}.
The {\sl Pixon} and {\sl Forward-Fit} algorithms yield compatible widths,
while those from {\sl MEM} are significantly larger, 
and those from {\sl MEM-Vis} are systematically narrower (probably
a side-effect of compromised image quality for time-variable sources).
The larger widths obtained from {\sl Clean} maps result from the convolution with the
{\sl Clean beam}. Apparently, {\sl MEM} does not converge to the best
solution when there are too many degrees of freedom.

The peak flux amplitude $a$ of an image map is roughly reciprocal to
the square of the source widths, i.e.m $a \propto
1/w^2$, because the flux per feature ($F \approx a w^2$) is well-conserved.
Because of this strong sensitivity to the
reconstructed source width, it is not unusual that the peak
flux differs up to an order of magnitude between different
image algorithms, depending on the convergence behavior of the
reconstruction algorithm. For imaging spectroscopy, therefore, it is important to
sum over an entire feature (such as a flare footpoint) rather
than to examine the flux of a source on a pixel by pixel basis.

\subsection{ 		Halo/Core Flux Ratios (\S A6 on CD-ROM)			}

We measure the flux ratio between a {\sl core (C)} region (inside a radius of 1.5 the resolution
of the finest grid) and the outer {\sl halo (H)} region. This H/C flux ratio reflects how many 
photons are spread in the image outside the real sources. It is found to be low for
{\sl Forward-fit, Pixon}, and {\sl Mem-Vis}, in the order of
a few percent, while {\sl Clean} shows a somewhat larger ratio of
$H/C\lapprox 30\%$ due to the convolution with the clean beam and the addition of clean residuals. 
{\sl MEM} exhibits a large noise ratio of $H/C\lapprox 160\%$, a clear indication
that the image algorithm does not converge properly in some cases, in particular
for small pixel sizes (1") and large field-of-views, which corresponds to a larger
degree of freedom in the image reconstruction.

\subsection{ 		Spatially Resolved Spectra (\S A7 on CD-ROM)			}

We fit a two-component (thermal plus non-thermal powerlaw) spectrum to the image
fluxes obtained in 8 different energy channels. Comparisons of the best-fit parameters
with those obtained with the spectral fitting software SPEX show agreement between
the non-thermal fluxes and power-law slopes within a few percents for all imaging
algorithms.
However, we emphasize that spatially-resolved spectra should only be extracted
from images by integrating over an entire source structure (or image feature),
never from single pixels (because of the amplitude-width uncertainty, see \S 2.5). 

\subsection{ 		Grid Selection (\S A8 on CD-ROM)			}

The grid selection determines the width of the reconstructed sources
in the case of unresolved sources. We found that the reconstructed source
width indeed scales (within a factor of 2) with the spatial
resolution of the finest used grid for most of the image
algorithms (except for {\sl MEM-Vis}, which over-resolves, and for
{\sl MEM}, which under-resolves the sources).

\subsection{ 		Pixel-size Comparisons (\S A9 on CD-ROM)			}

The image reconstruction should provide a solution that is independent of the map parameterization,
such as the field-of-view or pixel size, provided that the sources are resolved. Only in the case
of unresolved sources, the reproduced source width represents an upper limit specified by the
pixel size. We reconstruct the images with pixel sizes of 1", 2", and 4" and find indeed an
invariant solution for 4 algorithms, with a width that is related to the resolution of the
finest used grid. Only MEM-VIS shows a different behavior, yielding a source size that scales
with the pixel size rather than with the resolution of the finest used grid. 

\subsection{ 		Field-of-View Comparisons (\S A10 on CD-ROM)			}

The selected field-of-view is found not to affect the quality of the image reconstruction
for {\sl Forward-fit}, {\sl Pixon}, and {\sl Clean}. For {\sl MEM} and
{\sl MEM-Vis}, the quality of the image reconstruction seems to improve
for minimal field-of-views that closely encompass the sources, apparently 
because this minimizes the number of degrees of freedom in the image
reconstruction.

\subsection{ 		Time Interval Comparisons (\S A11 on CD-ROM)			}

Since the time interval is proportional to the number of photons used
in the image reconstruction, the quality and signal-to-noise ratio is
expected to improve for longer time intervals. We find that the signal-to-noise
ratio indeed improves with longer time intervals for {\sl Pixon},
{\sl Clean}, and {\sl MEM-Vis}. {\sl Forward-fit} is a noise-free model
by definition, so that a noise-to-signal ratio is not defined. For {\sl MEM}
we find no noise-to-signal improvement for longer time intervals, when the
number of degrees of freedom are too large (for small pixel sizes and
large field-of-views).

\section{		CONCLUSIONS				}

This study demonstrates that most available {\sl RHESSI} image
algorithms have a reasonable photometric accuracy for the total flux
contained in the images, but differ substantially in the geometry of
the reconstructed sources.  The results can be summarized as follows:
Three algorithms ({\sl Forward-Fit, Pixon, Clean}) converge to the
best solution within the constraints of the spatial resolution of the
finest used grids, within the expected uncertainty of the signal-to-noise 
dictated by photon statistics. The current implementation of the {\sl MEM} 
algorithm does not converge well for cases with a large number of degrees 
of freedom (for small pixel sizes and large field-of-views). The current
version of {\sl MEM-Vis} does not correct for time variability and thus
yields a compromised image quality for long-duration time intervals. 
Nevertheless, the user can produce consistent and high-quality
images with the {\sl MEM} and {\sl MEM-Vis} algorithms by exercising a
careful choice of image parameters.

\bigskip {\sl Acknowledgements:} We acknowledge constructive
and critical comments from David Smith and Richard Schwartz.
Support for this work was provided by the NASA SMEX grant
NAS5-98033 through University of California, Berkeley
(subcontract SA2241-26308PG).

\end{article}
\end{document}